\documentclass[aps,pra,twocolumn,groupedaddress,showpacs,floatfix]{revtex4}

\usepackage[dvips]{graphicx}
\usepackage{epsfig}
\usepackage{amssymb}
\usepackage{amsmath}
\usepackage{color}
\usepackage{hyperref}

\begin{document}


\title{Electrically tuned F\"orster resonances in collisions of NH$_3$ with Rydberg He atoms}


\author{V. Zhelyazkova and S. D. Hogan}

\affiliation{Department of Physics and Astronomy, University College London, Gower Street, London WC1E 6BT, U.K.}


\date{\today}

\begin{abstract} 
Effects of weak electric fields on resonant energy transfer between NH$_3$ in the X\,$^1$A$_1$ ground electronic state, and Rydberg He atoms in triplet states with principal quantum numbers $n = 36$\,--\,$41$ have been studied in a crossed beam apparatus. For these values of $n$, electric-dipole transitions between the Rydberg states that evolve adiabatically to the $|ns\rangle$ and $|np\rangle$ states in zero electric field can be tuned into resonance with the ground-state inversion transitions in NH$_3$ using electric fields, with energy transfer occurring via F\"orster resonance. In the experiments the Rydberg He atoms, traveling in pulsed supersonic beams, were prepared by resonant two-photon excitation from the metastable $1s2s\,^3S_1$ level and crossed an effusive beam of NH$_3$ before being detected by state-selective pulsed-electric-field ionization. The resonant-energy-transfer process was identified by monitoring changes in the ionization signal from the $|ns\rangle$ and $|np\rangle$ Rydberg states for each value of $n$. The electric field dependence of the experimental data is in good agreement with the results of calculations in which the resonant dipole-dipole coupling between the collision partners was accounted for. 
\end{abstract}


\maketitle

\section{Introduction}

Energy transfer between quantum systems coupled by resonant electric dipole interactions plays an important role in inter- and intra-molecular dynamics, particularly in excited electronic states~\cite{andrews99a,scholes03a}. These F\"orster resonance processes~\cite{perrin27a,forster46a,forster48a} are, for example, exploited in single molecule spectroscopy~\cite{ha96a}, and contribute to excitation transfer in light-harvesting complexes~\cite{cheng09a}. The electric dipole-dipole coupling, $V_{\mathrm{dd}}$, that gives rise to such energy transfer between two systems, A and B, has the form~\cite{gallagher92a}
\begin{eqnarray}
V_{\mathrm{dd}}(\vec{R}) &=& \frac{1}{4\pi\epsilon_0}\left[\frac{\vec{\mu}_{\mathrm{A}}\cdot\vec{\mu}_{\mathrm{B}}}{R^3} -  3\frac{(\vec{\mu}_{\mathrm{A}}\cdot\vec{R})(\vec{\mu}_{\mathrm{B}}\cdot\vec{R})}{R^5}\right],\label{eq:dd}
\end{eqnarray}
where $\vec{\mu}_{\mathrm{A}} = \langle \mathrm{f_A} | \hat{\vec{\mu}}_{\mathrm{A}} | \mathrm{i_A} \rangle$ and $\vec{\mu}_{\mathrm{B}} = \langle \mathrm{f_B} | \hat{\vec{\mu}}_{\mathrm{B}} | \mathrm{i_B} \rangle$ are the electric dipole transition moments between the initial, i, and final, f, internal quantum states of A and B, respectively, and $R = |\vec{R}|$ is the distance between the systems. The angular dependence of the scalar products in Eq.~(\ref{eq:dd}) reflects the sensitivity of the energy transfer process to the orientation of the dipoles, and hence the characteristics of their local environment. Consequently, $V_{\mathrm{dd}}$, and the rates of energy transfer, can be strongly modified by static, or time-dependent, electric and magnetic fields.

Because of (i) the large electric dipole moments of transitions between Rydberg states with high principal quantum number $n$, e.g., $|\langle np | \hat{\vec{\mu}} |ns\rangle| \geq1000\,ea_0$ for $n \geq 36$ in He, (ii) the tunability of Rydberg-Rydberg transition wavenumbers using electric or magnetic fields, and (iii) the control over the environment that can be achieved in gas-phase experiments, Rydberg atoms have been successfully exploited as model systems with which to study F\"orster resonance energy transfer~\cite{gallagher92a,gallagher08a,richards16a, yakshina16a, maineult16a}. For example, resonant interactions between pairs of neutral atoms of a single species have been used to blockade laser photoexcitation~\cite{vogt06a} and generate pairwise entanglement~\cite{wilk10a}, realize strong optical nonlinearities in cold atomic gases~\cite{firstenberg16a} to demonstrate single photon optical transistors~\cite{tiarks14a,gorniaczyk16a}, and investigate energy transfer within tailored arrangements of atoms~\cite{ravets14a,gunter13a}.

In addition to studies of resonant energy transfer between atoms in Rydberg states, experiments have also been reported in which resonant depopulation and ionization of Rydberg atoms in collisions with polar molecules were observed in the absence of external fields~\cite{smith78a,petitjean84a,petitjean86a,ling93a}. More recently, in the context of the development of methods for the preparation of ultracold polar molecules it has been proposed that resonant interactions with cold Rydberg gases could be exploited to cool the molecules~\cite{huber12a,zhao12a}. To implement schemes of this kind it is essential to perform Rydberg-state-resolved studies of F\"orster resonances in these hybrid atom-molecule systems to identify and characterize the effects of external fields on these interactions. From a more general perspective, such measurements also offer new opportunities for investigations of the role of coherences between electronic, vibrational and rotational degrees of freedom (see, e.g., Refs.~\cite{scholes15a,scholes10a} and references therein) and environmental effects in molecular resonance energy transfer, and studies of chemical dynamics in which long-range dipolar interactions can be exploited to regulate access to short-range Penning ionization processes~\cite{jankunas16a} at low temperature.

In this paper we report the results of experiments in which we observe F\"orster resonance energy transfer from the inversion sublevels of the X\,$^1$A$_1$ ground electronic state of NH$_3$, to He atoms in selected triplet Rydberg states with values of $n$ from~36 to~41. Using electric fields up to 12~V\,cm$^{-1}$, electric-dipole transitions between the states that evolve adiabatically to the $|ns\rangle$ and $|np\rangle$ Rydberg states in zero electric field were tuned through resonance with the NH$_3$ inversion spectrum, modifying the energy transfer rates. The interpretation of the experimental data is aided by comparison with the results of calculations of the electric-field dependence of the energy transfer process. 

\section{Experiment}

\begin{figure}
\begin{center}
\includegraphics[width = 0.46\textwidth, angle = 0, clip=]{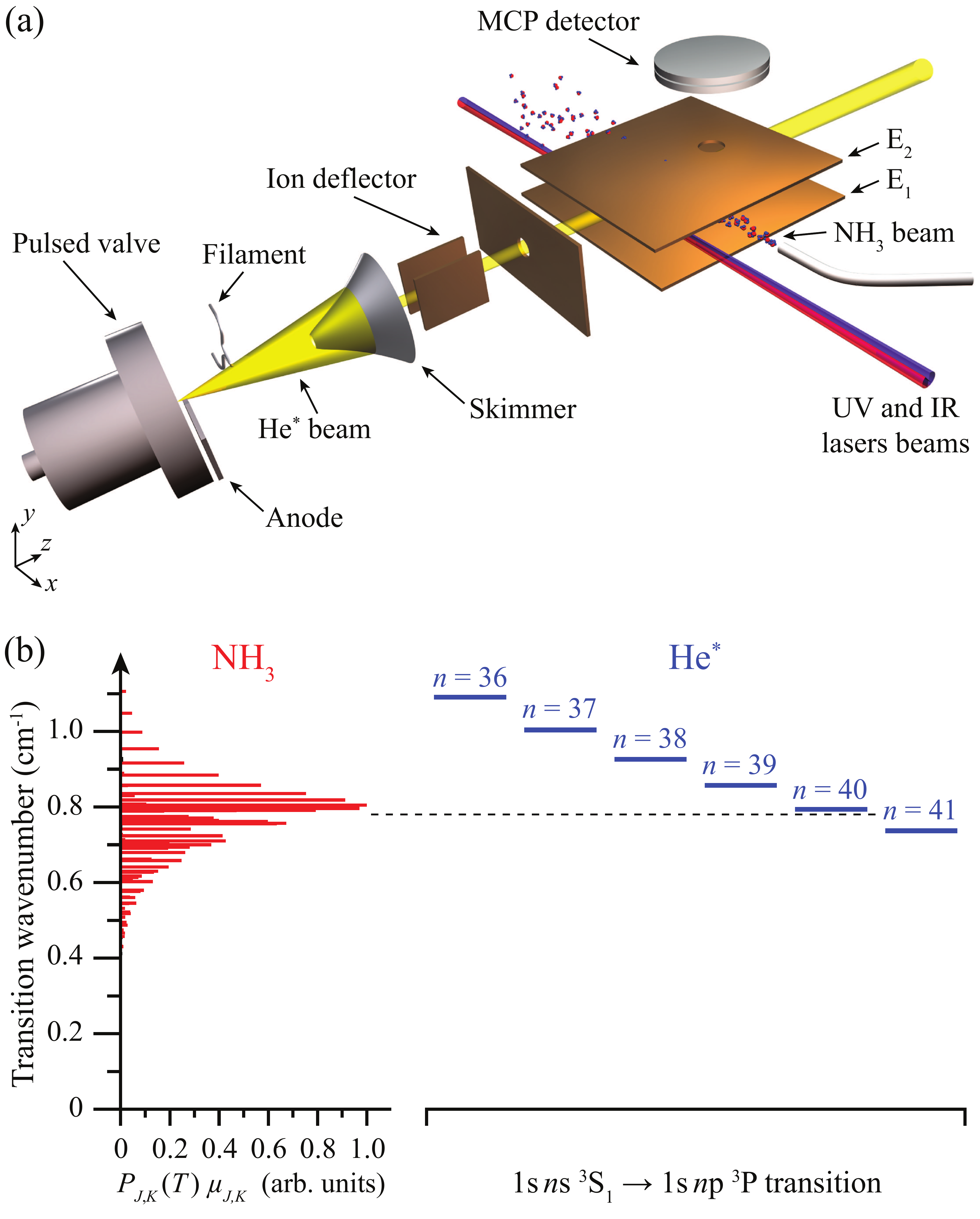}
\caption{(a) Schematic diagram of the experimental apparatus (not to scale). (b) Comparison of the transition-dipole-moment-weighted ground-state inversion spectrum of NH$_3$ at 300~K with the field-free wave numbers of $1sns\,^3S_1\rightarrow1snp\,^3P_J$ transitions in He.}
\label{fig1}
\end{center}
\end{figure}

An overview of the experimental apparatus is presented in Fig.~\ref{fig1}(a). Helium atoms traveling in pulsed supersonic beams (mean longitudinal speed 2000~m\,s$^{-1}$) were prepared in the metastable $1s2s\,^3S_1$ level in a dc electric discharge at the exit of a pulsed valve~\cite{halfmann00}. After collimation and deflection of stray ions, the He beam was intersected by UV and IR laser radiation between two parallel 70~mm$\times$70~mm copper electrodes, E$_1$ and E$_2$ in Fig.~\ref{fig1}(a), separated by 8.4~mm. The UV laser was frequency stabilized to drive the $1s2s\ ^3S_1\rightarrow 1s3p\ ^3P_2$ transition at 25\,708.588~cm$^{-1}$ ($\equiv388.975$~nm). The IR laser was tuned in the range from 12\,660 to 12\,681~cm$^{-1}$ ($\equiv789.89$ to 788.58~nm) to subsequently excite $1sns\;^3S_1$ levels with $n=36$ -- $41$. After photoexcitation, the typical number density of excited Rydberg atoms was $\sim10^8$~cm$^{-3}$~\cite{zhelyazkova16b}.

Rydberg state photoexcitation was performed in zero electric field for 3~$\mu$s. After this time, the field was switched (rise time 200~ns) to a selected value, $F_{\mathrm{int}}$, by applying appropriate potentials to E$_2$, as the Rydberg atoms crossed an effusive beam of NH$_3$ emanating from a 1-mm-diameter tube maintained at 1.8~mbar and 300~K (mean speed $\sim720$~m\,s$^{-1}$; number density $\sim2\times10^9$~cm$^{-3}$~\cite{scholes88a}). After an interaction time of 5~$\mu$s, $F_{\mathrm{int}}$ was switched back to zero before the Rydberg atoms were detected by ramped electric field ionization upon applying pulsed potentials of up to $V_{\mathrm{ion}}\simeq-300$~V, rising with an RC-time-constant of 1.0~$\mu$s, to E$_1$. The ionized electrons were accelerated through a 4-mm-diameter hole in E$_2$ to a microchannel plate (MCP) detector. In this process, the Rydberg atoms in the center of the excited ensemble spent approximately equal periods of time (5~$\mu$s) in the interaction field $F_{\mathrm{int}}$, and in zero electric field. 

\section{Experimental results}

\begin{figure}
\begin{center}
\includegraphics[width=0.48\textwidth, angle = 0, clip=]{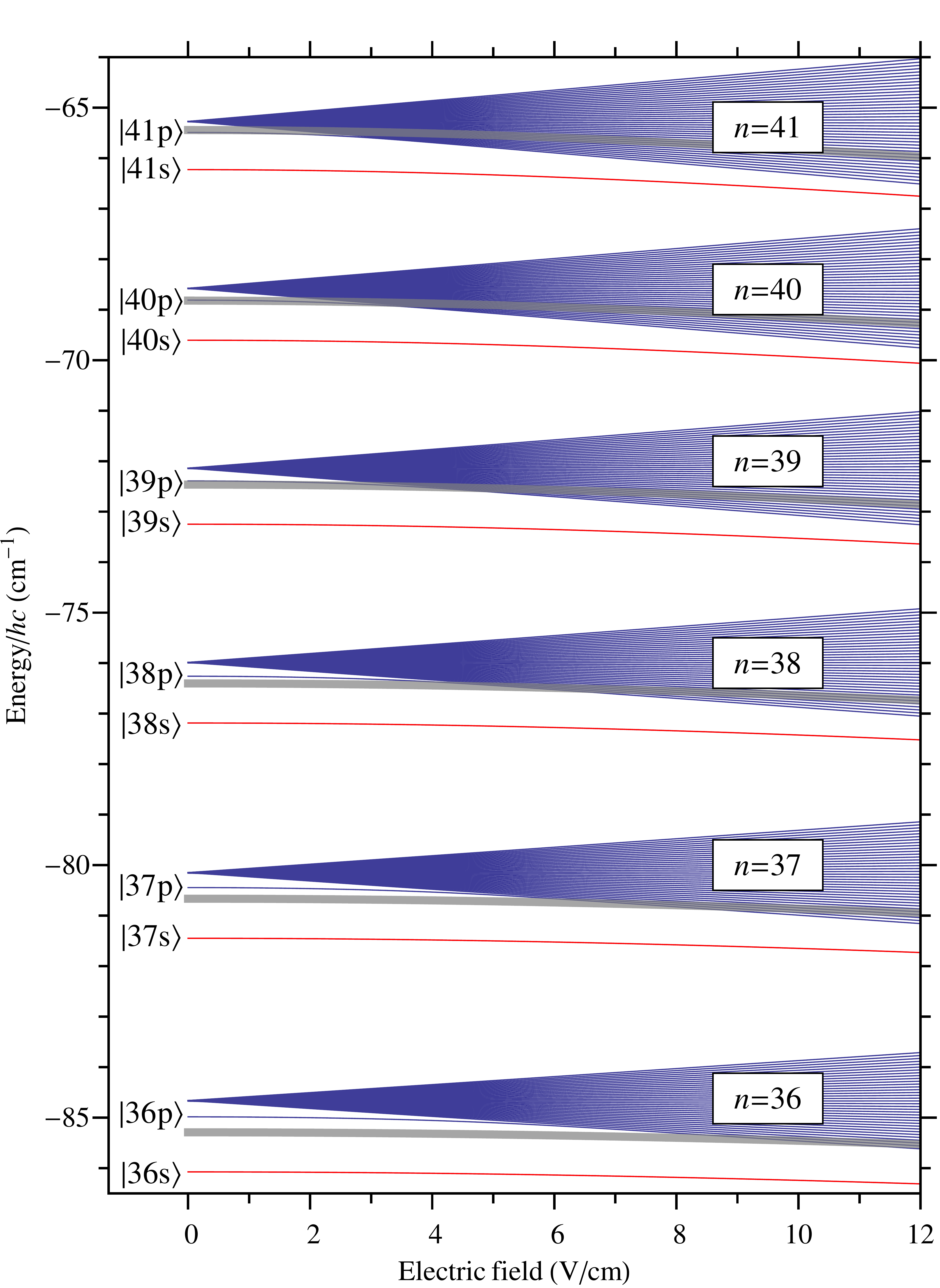}
\caption{Energy level structure of the $m_{\ell}=0$ triplet Rydberg states in He with $n=36$\,--\,$41$. The $|ns\rangle$ levels photoexcited in the experiments are indicated in red. Shown as the gray 
shaded curves are the wavenumbers of each $|ns'\rangle$ level offset by the centroid inversion wavenumber in NH$_3$ ($\sim0.78$~cm$^{-1}$).}
\label{fig2}
\end{center}
\end{figure}

The $|ns\rangle$ and $|np\rangle$ triplet Rydberg states in He (referring to the $1sns\,^3S_1$ and $1snp\,^3P_J$ levels in zero field, respectively) are separated by 0.73 -- 1.1~cm$^{-1}$ for $n=36$\,--\,$41$, while the inversion transitions in the X\,$^1$A$_1$ state of NH$_3$ lie at $\sim0.78\pm0.2$~cm$^{-1}$ at 300~K. The $|ns\rangle\rightarrow|np\rangle$ transition is therefore resonant with the NH$_3$ inversion spectrum at $n\simeq40$  [see Fig.~\ref{fig1}(b)]. The energy level structure of the $m_{\ell}=0$ triplet Rydberg states of He is displayed in Fig.~\ref{fig2} for electric fields up to 12~V\,cm$^{-1}$. Because of their non-zero quantum defects, e.g., $\delta_{36s}$=0.29669 and $\delta_{36p}$=0.06835~\cite{drake99}, the $|ns\rangle$ and $|np\rangle$ states at each value of $n$ lie lower in energy in zero field than the higher angular momentum states. In increasing fields, the wavenumber intervals between these two states reduces as they evolve into the $|ns'\rangle$ and $|np'\rangle$ $\ell-$mixed Stark states. The gray shaded curves in Fig.~\ref{fig2}, shifted in wavenumber from each $|ns'\rangle$ state by the centroid NH$_3$ inversion wavenumber ($\sim$0.78~cm$^{-1}$), show that electric fields in the range encompassed in the figure can be used to tune the two systems through resonance.   

\begin{figure}
\begin{center}
\includegraphics[width=0.48\textwidth, angle = 0, clip=]{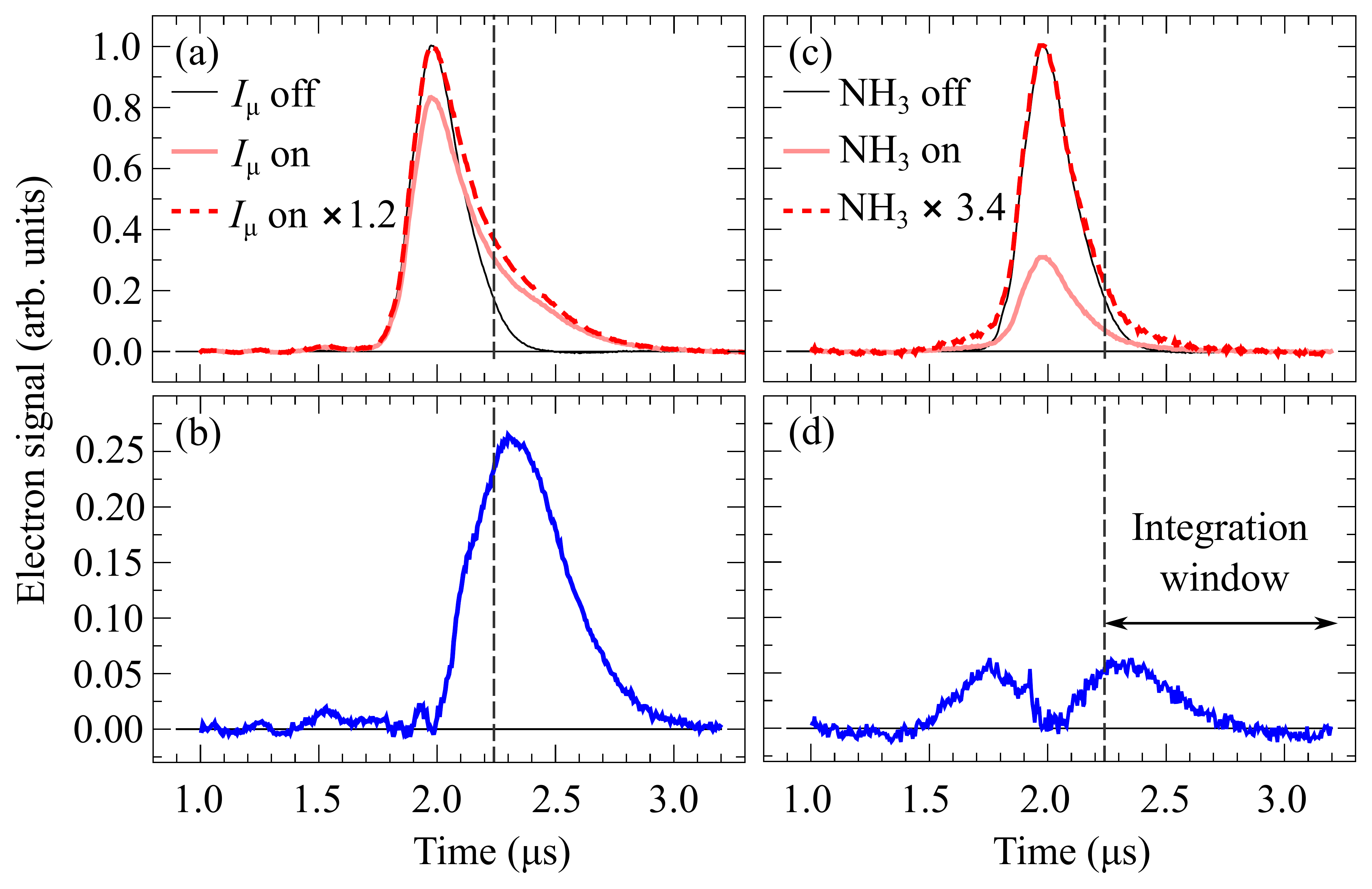}
\caption{Electric field ionization profiles at $n=38$ in He demonstrating the effect of microwave transfer of population from the $|38s\rangle$ state to the $|38p\rangle$ state [(a) and (b)], and collisions with NH$_3$ [(c) and (d)]. In (a) and (c) the unperturbed $|38s\rangle$ signal is indicated by the thin black curve, and the un-normalized (normalized) profiles recorded following microwave transfer or collisions are indicated by the thick pink (dashed red) curves. The dashed vertical line in each panel marks the start of the integration window used to identify the $|np\rangle$ signal in Fig.~\ref{fig4}.}
\label{fig3}
\end{center}
\end{figure}

Because the $|ns\rangle$ and $|np\rangle$ Rydberg states ionize in different electric fields, energy transfer in collisions with NH$_3$ can be identified from the electron time-of-flight profiles in the time-dependent ionization fields. The slight asymmetry in this profile results from the transient response of the RC charging circuit used to generate the ramped ionization electric field. An example of such a ramped field ionization profile for the $|38s\rangle$ state is displayed in Fig.~\ref{fig3}(a) (thin black curve). To identify changes in this profile when the $|38p\rangle$ state is populated, a reference measurement was made in which population was transferred to the $|38p\rangle$ state with $|m_{\ell}|=1$ by a pulse of microwave radiation with no NH$_3$ present [thick pink curve in Fig.~\ref{fig3}(a)]. Following normalization of these data recorded with the microwaves on [dashed red curve in Fig.~\ref{fig3}(a)] and subtraction of the profile for the $|38s\rangle$ state alone, the contribution from the $|38p\rangle$ state to the signal is seen [Fig.~\ref{fig3}(b)]. In the time-dependent ionization fields used in the experiments the $|np\rangle$ states with $|m_{\ell}|=1$ ionize quasi diabatically and therefore at higher fields than the $|ns\rangle$ states which ionize adiabatically. The $|np\rangle$ states with $m_{\ell}=0$ also ionize adiabatically, but in fields slightly lower than those of the $|ns\rangle$ states because of their smaller quantum defects. Consequently, complete $|m_{\ell}|$-state selective detection was achieved. 

\begin{figure}
\begin{center}
\includegraphics[width=0.48\textwidth, angle = 0, clip=]{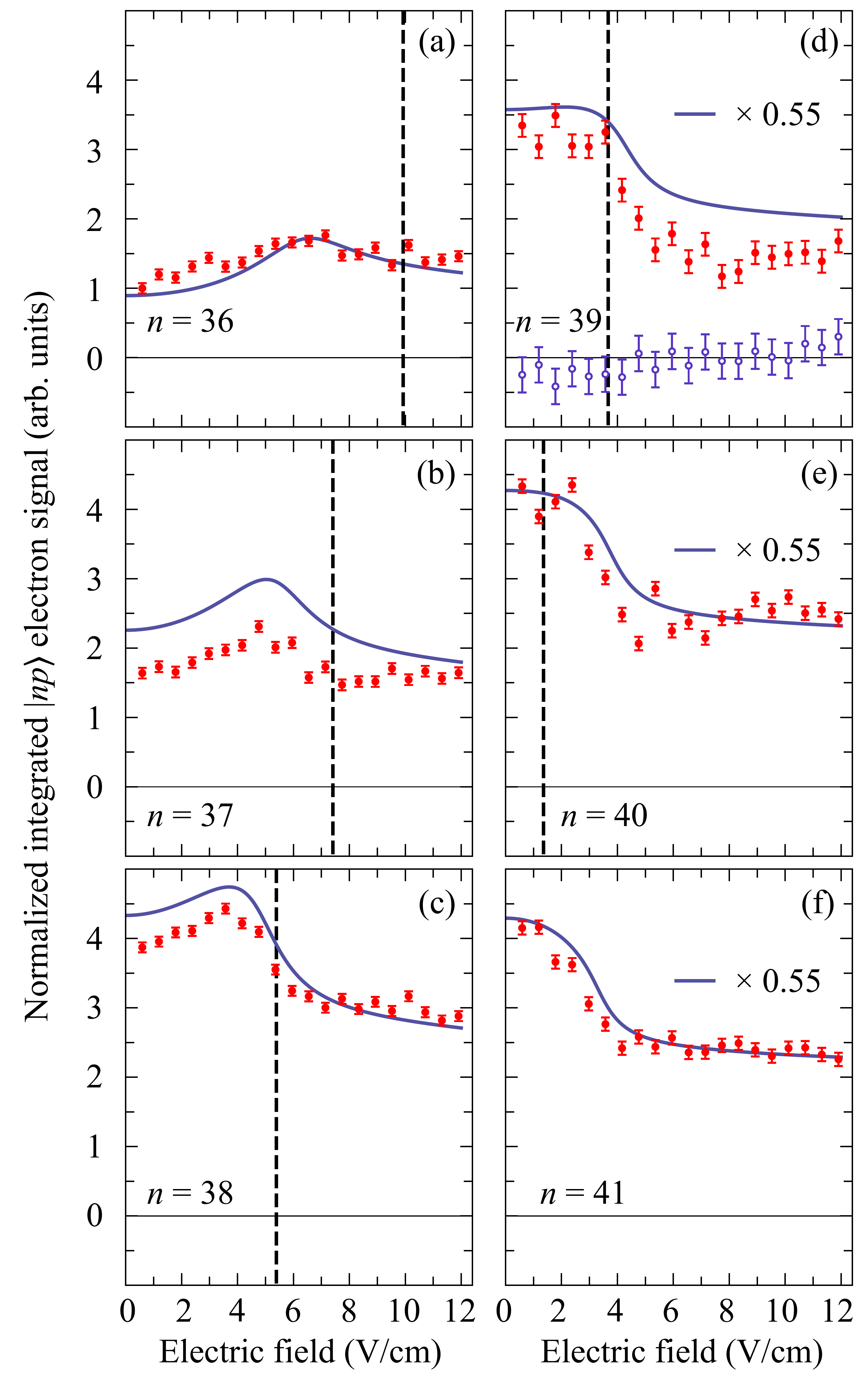}
\caption{(a)-(f) Integrated $|np\rangle$ electron signal recorded following collisions of NH$_3$ with He atoms prepared in $|ns\rangle$ states with $n=36$\,--\,$41$. All data sets are normalized relative to the first data point in (a). The dashed vertical lines indicate the electric field at each value of $n$ for which the centroid inversion splitting in NH$_3$ is resonant with the $|ns'\rangle\rightarrow |np'\rangle$ transition in He. The continuous curves represent the normalized dependence of the calculated $|np\rangle$-electron signal in the electric field. In (d)-(f) the results of the calculations are scaled by a factor of 0.55. A reference measurement with the NH$_3$ beam replaced with a beam of NO is indicated by the open circles in (d).}
\label{fig4}
\end{center}
\end{figure}

Using this $|38p\rangle$ signal as a guide, the effects of collisions with the beam of NH$_3$, in the absence of the microwaves were determined. The ionization profile of the $|38s\rangle$ state with $F_{\mathrm{int}}=0.6$~V\,cm$^{-1}$ and NH$_3$ present is displayed in Fig.~\ref{fig3}(c) (thick pink curve). The total electron signal is significantly reduced following collisions with the NH$_3$ as a result of $n$-changing associated with rotational energy transfer, and collisional ionization~\cite{petitjean86a}.  When normalized to the intensity maximum of the pure $|38s\rangle$ signal, the state-changing effects of the collisions can be seen, with some population detected in the ionization time window associated with the $|38p\rangle$ state. Similar profiles for $n=36$\,--\,$41$ all displayed population of the $|np\rangle$ state at ionization times beyond $2.24$~$\mu$s when $V_{\mathrm{ion}}$ was set so that the $|ns\rangle$-electron signal was detected at the same time of flight for each value of $n$. Having identified the time window corresponding to the ionization of atoms in the $|np\rangle$ states, the effects of the electric fields on the energy transfer process were  investigated. This was done by integrating the electron signal in the $|np\rangle$ detection window [see Fig.~\ref{fig3}(d)] for $F_{\mathrm{int}} = 0.6$\,--$12$~V\,cm$^{-1}$, with the results displayed in Fig.~\ref{fig4}. 
     
For the lower values of $n$ [e.g., $n=36-38$, in Fig.~\ref{fig4}(a-c)] the zero-field $|ns\rangle\rightarrow|np\rangle$ transition lies above the centroid transition wavenumber of the NH$_3$ inversion spectrum [see Fig.~\ref{fig1}(b)]. When $F_{\mathrm{int}}$ is increased, the reduction in the wavenumber interval between the $|ns'\rangle$ and $|np'\rangle$ states brings the energy transfer channel into resonance and causes an increase in the $|np\rangle$-electron signal observed up to $F_{\mathrm{int}}\simeq7$, 5, and 4~V\,cm$^{-1}$ in Fig.~\ref{fig4}(a), (b) and (c), respectively. For values of $n$ between 39 and 41 [Fig.~\ref{fig4}(d-f)], there is sufficient overlap between the zero-field $|ns\rangle\rightarrow |np\rangle$ transition wavenumber and the inversion spectrum for resonance energy transfer to occur even in the weakest fields investigated. In these measurements, and in fields exceeding 8~V\,cm$^{-1}$ for $n<39$, the $|np\rangle$-electron signal reduces as $F_{\mathrm{int}}$ is increased. This reflects the shift away from resonance of the energy transfer process. 

To confirm the origin of the resonance features in the data in Fig.~\ref{fig4}, the NH$_3$ beam was replaced with an effusive beam of NO emanating from the same source. Since there are no electric-dipole transitions in the X\,$^2\Pi$ ground state of NO in the spectral region close to 1~cm$^{-1}$, no resonant transfer to the $|np\rangle$ state is expected for the fields in Fig.~\ref{fig4}. This is confirmed in the data recorded at $n=39$ and presented in Fig.~\ref{fig4}(d) (open circles).

\section{Calculations}

To obtain further insight into the F\"orster resonance energy transfer process, and the dependence of the experimental data in Fig.~\ref{fig4} on $F_{\mathrm{int}}$, energy-transfer cross sections were calculated following the approach employed to treat collisions between pairs of Rydberg atoms in Ref.~\cite{gallagher92a}. These calculated cross sections were used to determine the resonance widths and, when combined with the Stark shifts of the $|ns'\rangle$ and $|np'\rangle$ Rydberg states, the dependence of the $|np\rangle$-electron signal on $F_{\mathrm{int}}$. 

For a constant collision speed, the cross section for resonant energy transfer, $\sigma_{J,K}(F_{\mathrm{int}})$, from each rotational state $|J,K\rangle$ in NH$_3$, where $J$ is the total angular momentum quantum number and $K$ is the projection of $\vec{J}$ onto the symmetry axis of the molecule, to a He Rydberg atom is 
\begin{eqnarray}
\sigma_{J,K}(F_{\mathrm{int}}) &\propto& \mu_{ns',np'}(F_{\mathrm{int}})\,\mu_{J,K},\label{eq:crosssec}
\end{eqnarray}
where $\mu_{ns',np'}(F_{\mathrm{int}})=|\langle np'| \hat{\vec{\mu}} |ns'\rangle|$ is the electric dipole transition moment between the $|ns'\rangle$ and $|np'\rangle$ states, and
\begin{eqnarray}
\mu_{J,K} = \sqrt{\frac{\mu_0^2\,K^2}{J(J+1)}},
\end{eqnarray}
is the transition dipole moment associated with the inversion of the NH$_3$ molecule in the $|J,K\rangle$ state, with $\mu_0=1.468$~D~\cite{townes55a}. The Rydberg-Rydberg transition dipole moments were calculated from the eigenvectors of the Hamiltonian matrix describing the interaction of the He atom with the electric field~\cite{zimmerman79a}. In the experiments the NH$_3$ molecules were not aligned or oriented. To account for this in the calculations, $V_{\mathrm{dd}}$ in Eq.~(\ref{eq:dd}) was averaged over all angles and since the Stark energy shift of the NH$_3$ ground state in fields up to 12~V\,cm$^{-1}$ is $<0.001$~cm$^{-1}$, it was neglected. Under these conditions, in zero electric field the calculated energy transfer cross section for $n=40$ is $\sim2\times10^{-11}$~cm$^2$. Assuming a pseudo-first-order kinetic model, this corresponds, under the experimental conditions, to an energy transfer rate of $\sim10^4$~s$^{-1}$, or a zero-field transition probability in 10$~\mu$s interaction time of $\sim0.1$. This is on the order of that observed in Fig.~\ref{fig3}(c-d). The measured energy transfer rates depend on the Boltzmann factors, $P_{J,K}(T)$, reflecting the relative population of each $|J,K\rangle$ state~\cite{townes55a}. These quantities, and the wavenumbers of the inversion transitions were calculated using the rotational constants $B = 9.44$~cm$^{-1}$ and $C = 6.20$~cm$^{-1}$~\cite{townes55a,costain51}. The products of $P_{J,K}(T)$ and $\mu_{J,K}$, i.e., the contribution from each $|J,K\rangle$ state to the energy transfer rate, are displayed in Fig.~\ref{fig1}(b).   

The measured $|np\rangle$-electron signal associated with the energy transfer process from each $|J,K\rangle$ state is proportional to $P_{J,K}(T)\,\sigma_{J,K}(F_{\mathrm{int}})$ and dependent upon the detuning from resonance. Assuming a Gaussian form, with a full-width-at-half-maximum of $\Delta E_{\mathrm{FWHM}}/hc\propto1/\sqrt{\sigma_{J,K}(F_{\mathrm{int}})}$, for the dependence on the detuning~\cite{gallagher92a}, the continuous curves in Fig.~\ref{fig4} were obtained with one global fit parameter -- the constant of proportionality associated with the resonance width. This was found to be $18$~D~cm$^{-1}$ resulting in typical atom-molecule interaction times, e.g., at $n=40$ in zero electric field, of 0.2~ns and hence resonance widths of $\Delta E_{\mathrm{FWHM}}/hc\simeq0.17$~cm$^{-1}$ ($\Delta E_{\mathrm{FWHM}}/h\simeq5$~GHz). 

The results of the calculations displayed in Fig.~\ref{fig4} account for the equal periods of 5~$\mu$s that the atoms spend in zero electric field and in $F_{\mathrm{int}}$, and are in good agreement with the measured dependence of the energy transfer process on $F_{\mathrm{int}}$. It is noticeable that the fields for which maximal energy transfer is observed at each value of $n$ are lower than those for which the Rydberg-Rydberg transition wavenumber coincides exactly with the centroid of the NH$_3$ inversion spectrum (dashed vertical lines in Fig.~\ref{fig4}). This shift arises from the dependence of $\mu_{ns',np'}(F_{\mathrm{int}})$ on $F_{\mathrm{int}}$. For each value of $n$, $\mu_{ns',np'}(F_{\mathrm{int}})$ is maximal when $F_{\mathrm{int}}=0$~V\,cm$^{-1}$. As the $|ns\rangle$ and $|np\rangle$ states mix in the field, $\mu_{ns',np'}(F_{\mathrm{int}})$ decreases. Consequently, for each value of $n$, the dipole-dipole coupling between the collision partners is largest in electric fields below those that satisfy the resonance condition, and fall off as the field is tuned through resonance. For the higher values of $n$, Fig.~\ref{fig4}(d)-(f), the measured population transfer to the $|np\rangle$ states is lower than expected when compared to the observations at $n=36$. This may be a result of additional energy transfer channels opening as the value of $n$ increases~\cite{smith78a}, or an effect of individual atoms undergoing multiple collisions, with a small contribution from the $n$-dependence of the Rydberg state fluorescence lifetimes which range from 36~$\mu$s (56~$\mu$s) to 53~$\mu$s (83~$\mu$s) for the $|36s\rangle$ ($|36p\rangle$) to the $|41s\rangle$ ($|41p\rangle$) states in zero field. 

\section{Conclusion}

The studies of effects of electric fields on F\"orster resonance energy transfer between NH$_3$ and Rydberg He atoms reported here open opportunities for investigations of the contributions of collision energy, particle orientation and coherence in the energy transfer process. By using beams of strongly interacting Rydberg atoms with large static electric dipole moments~\cite{zhelyazkova16b}, it will be possible to investigate many-body effects in carefully controlled environments. To achieve maximal control over the collisional partners it will be desirable to employ guided or decelerated molecular beams~\cite{bethlem99a,janukas15a,vandemeerakker12a}. These techniques, combined with the methods of Rydberg-Stark deceleration~\cite{hogan16a}, are well-suited to low-energy collision experiments~\cite{onvlee16a,allmendinger16a}.

\emph{Note added in proof}. Recently, it has been suggested that resonant energy transfer between polar ground state molecules and Rydberg atoms could be exploited for non-destructive detection of cold molecules~\cite{zeppenfeld16a}.

\begin{acknowledgments}
This work is supported by the Engineering and Physical Sciences Research Council under Grant No. EP/L019620/1, and the European Research Council (ERC) under the European Union's Horizon 2020 research and innovation program (Grant Agreement No 683341).
\end{acknowledgments}

\end{document}